\documentclass[prb]{revtex4}
\newcommand{\be}{\begin{equation}}
\newcommand{\ee}{\end{equation}}
\newcommand{\ba}{\begin{eqnarray}}
\newcommand{\ea}{\end{eqnarray}}
\usepackage{graphicx}
\usepackage{dcolumn}
\usepackage{bm}
\usepackage{amsfonts}

\begin{document}

\preprint{}

\title{Elasticity of soft particles and colloids near the jamming threshold}
\author{ Matthieu Wyart}

\affiliation{School of Engineering and Applied Sciences, Harvard University, 29 Oxford Street, Cambridge, MA 02138\\
Lewis-Sigler Institute, Princeton University, Princeton, NJ 08544-1014}
\date{\today}

\begin{abstract}

Assemblies of purely repulsive and frictionless particles, display very curious elastic properties near the jamming transition where the overlap between particles vanish. 
Although such systems do not contain the long and cross-linked polymeric chains characterizing  a rubber, they behave macroscopically in a similar way: the shear modulus $G$ can become negligible in comparison with the bulk modulus $B$, as observed in emulsions \cite{mason}. Numerics \cite{J,mason} have established a scaling relationship between the ratio of elastic moduli and the compression in systems of athermal compressed elastic particles. After reviewing recent theoretical results  on the microscopic structure of such packings (in particular the coordination number $z$, the average number of contact per particle) I will  propose an explanation for these observations, and explain why the arguments apply both to soft athermal particles  and to colloids where elasticity has an entropic nature.

\end{abstract}
\maketitle{}
\section{Introduction}

Crystalline lattices are invariant under translation, implying that vibrational modes are plane waves. From those excitations, one can build a theory of energy transport and a theory of elasticity.
In amorphous solids, this symmetry breaks down. Although at large length scales a continuous (and translationary invariant)  description is a good approximation, this fails at small length scales, where the disorder has strong effects.
At these scales, various properties of amorphous solids, such as energy transport, low-frequency excitations in glasses or force propagation in granular matter, 
are not yet satisfyingly understood, and are active fields of research. One inherent difficulty in the study of these phenomena is that the length scales at play are typically moderate, of the order of ten particle sizes or less. 
This makes it harder to test  and distinguish clearly the consequences of different theories. Finding a system where length scales can be large and controlled may therefore be extremely useful. 

Emulsion experiments \cite{mason}, followed with theoretical arguments  \cite{shlomon,Tom1,Tom2,moukarzel} suggested that the ``jamming transition" where repulsive, short-range particles are just in contact corresponds to a critical point.  This idea was latter substantiated by the findings that the elastic moduli \cite{ J}, the vibrational spectrum \cite{J}, the microscopic structure \cite{durian, J} and force propagation \cite{wouter} display scaling behavior near the jamming threshold. At that point, although the system is amorphous and isotropic, it cannot be described as a continuous elastic body on any length scale \cite{matthieu1,wouter}. Because the strong effects of disorder occurs already at large length scales near this critical point, this model system is a lens allowing to probe in a stringent manner the properties of amorphous solids, and the effects of disorder. It has enabled to build and test \cite{matthieu1,these} a theory for the excess low-vibrational modes (the so-called Boson Peak) found in amorphous solids, which applies as well to covalent glasses \cite{these} and to model systems of attractive glasses \cite{these,ning}. This line of thought also permits to relate microscopic structure and some aspects of the dynamics near the glass transition of hard spheres \cite{brito}, despite the fact that the glass transition occurs empirically at a packing fraction significantly smaller than those of jammed configurations. 

In what follows I will focus on the elastic moduli, as micro-gels are potentially a good system to vary the packing fraction around the jamming threshold  to test predictions on elasticity. Near the jamming threshold it is found numerically that both for particles interacting with a harmonic potential (used to model emulsions) or a Hertzian potential (used to model elastic particles) \cite{J}, the system is ``almost" a liquid, its shear modulus becomes negligible in comparison with the bulk modulus:
\be
\label{1}
\frac{G}{B}\sim (p/B)^{1/2}
\ee
where $p$ is the pressure.  In a gel, two different phenomena govern the shear and the bulk modulus: the elasticity of the polymeric network and the compressibility of the solvent, respectively. This can cause the two elastic moduli to be very different. Nevertheless this behavior is unusual for a solid made of identical particles interacting with a radial interaction, where shear and bulk modulus are both induced by local interactions, and are generally comparable in amplitude. The theoretical argument, presented in Section \ref{s2}, will explain this observation  and unravel a peculiar aspect of the elasticity near the jamming threshold:  the elastic moduli can depend enormously on the stress applied on the system before the response is measured, the so-called pre-stress.  Before discussing this argument, I will start by reviewing recent results on the geometry of the packings near the jamming threshold.

\section{Structure and mechanical stability}
\label{s1}

Amorphous solids are typically out-of-equilibrium, 
and their structures a priori depend on their history. It may thus seem hard to infer their structure without a detailed description of the way they were made. 
There is, nevertheless, a limiting case that turns out to be conceptually important: when these systems are prepared via a very rapid quench from a fluid phase. In this situation, 
we expect that as soon as the liquid finds some meta-stable states, it remains in those states. For slower quenches the system will depart from such states, but there exists evidences that this effect is small,
at least for hard particles and for quenches as slow as what is typically achieved numerically \cite{brito}.  Thus we are particularly interested in configurations that are marginally stable, i.e mechanically stable, but which are very close to yield. 
What are those configurations for an assembly of elastic particles?

Maxwell \cite{max} studied mechanical stability in the context of engineering structures, and he found out that the key parameter is the coordination $z$, the average number of interactions per particle.
For example, for a network of point particle connected via springs, he showed that stability requires $z\geq z_c=2d$, where $d$ is the spatial dimension of the system. His argument goes as follows: 
consider a set of $N$ points interacting with $N_c$ springs at rest of stiffness $k$. The  expansion
for the energy may be written: 
\be 
\label{10} 
\delta E=\sum_{\langle ij\rangle} \frac{k}{2} [(\delta {\vec
R_i}-\delta {\vec R_j})\cdot {\vec n_{ij}}]^2 +o(\delta R^2)
\ee 
where the sum is made over all springs, ${\vec n_{ij}}$ is the unit
vector going from $i$ to $j$, and $\delta {\vec R_i}$ is
the displacement of particle $i$. A system is floppy, {\it i.e.} not mechanically stable,  if it
can be deformed without energy cost, that is if there is a displacement field for which $\delta E=0$, or
equivalently $(\delta {\vec R_i}-\delta {\vec R_j})\cdot
{\vec n_{ij}}=0 \ \forall ij$. If the spatial dimension is
$d$, this linear system has $Nd$ degrees of freedom and
$N_c\equiv Nz/2$ equations, and therefore there are always non-trivial
solutions if $Nd> N_c$, that is if $z<2d\equiv z_c$. To be mechanically stable a network must therefore have $z\geq2d$. 

\begin{figure}[htbp]
\begin{center}
\rotatebox{0}{\resizebox{4.5cm}{!}{\includegraphics{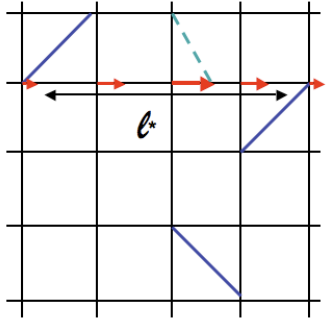}}}
\caption{Square lattice with a density per particle $\delta z$ of additional diagonal springs, represented in blue.  $l^*\sim 1/\delta z$ is the typical (dimensionless) distance of the segments contained between two diagonal springs on a given row or column.  The red arrows represent the longitudinal  mode of wavelength $\sim l^*$ of such a segment: $\delta {\vec R_i}\propto  \sin (\pi i  /l^*) {\vec e_x}$, following  the notation introduced in the text.  The  dashed line exemplifies the deformation of a spring transverse and directly connected to the segment considered, it is elongated by the longitudinal vibration of this segment.  When the pressure is positive and contacts are under compression, this elongation lowers the energy those springs contain. This leads to an instability when  $\delta z$ becomes smaller than a quantity proportional to the square root of the contact strain, of order $p/B$. }
\label{f1}
\end{center}
\vspace{-0.5cm}
\end{figure}

It turns out that under compression the criterion of rigidity becomes more demanding. 
Here we shall derive this criterion in a simple model, as it yields the correct and more general result. The derivation for a generic amorphous packing is made in \cite{matthieu2}. 
Consider a square lattice made of springs of rest length $l_0$, which defines our unit length. It just satisfies the Maxwell criterion, since $z=4$. Now, we add randomly a density $\delta z$ of springs connecting second neighbors, represented in blue in Fig(\ref{f1}), such that the coordination is $z=z_c+\delta z$. We add them in a rather homogeneous manner, 
so that there are not large regions without blue springs. The typical  distance between two blue springs in a given row or column is of order $l_0/\delta z$. Dividing this length by the mesh size $l_0$ we define the dimensionless number $l^*= 1/\delta z$ . 

How much pressure can this system sustain before collapsing? 
For a system to be mechanically stable, all  collective displacements need to have a positive energetic cost. 
It turns out that the first modes to collapse are of the type of the red displacement mode represented in Fig(\ref{f1}): they correspond to the longitudinal  mode of wavelength $l_0 l^*$ of a segment of springs contained between two blue, diagonal springs.  These modes have a displacement field of  the form $\delta {\vec R_i}=2 X \sin (\pi i  /l^*)/\sqrt{l^*} {\vec e_x}$, where $i$ labels the particles along a segment and runs between 0 and $l^*$, ${\vec e_x}$ is the unit vector in the direction of the line, and $X$ is the amplitude of the mode, $X=1$ for a normalized mode. In the absence of pressure $p$, the energy  of this mode comes only from the springs of the segment. In the limit of large $l^*$ a Taylor expansion of the displacement in Eq.(\ref{10}) gives the well-know result for the energy of the lowest-frequency mode of a line of springs $\delta E \sim k X^2/l^*{}^2 $. When $p>0$, all the springs now carry a force $f\sim p l_0$. The energy expansion contains other terms not indicated in Eq.(\ref{10}) \cite{matthieu2}, whose effect can be estimated quantitatively as follows. When particles are displaced along a longitudinal mode such as the one represented in Fig.(\ref{f1}), the force of each  spring directly connected and transverse to the segment considered ( see Fig(\ref{1}) ) now produces a work equal to $f$ times the elongation of the spring. This elongation is simply $\delta {\vec R_i}^2/ l_0$ following Pythagoras' theorem. Summing on all the springs transverse to the segment leads to a work of order $ f X^2/l_0\approx p X^2$. This gives finally for the energy of the mode $\delta E \sim k X^2/l^*{}^2  -p X^2$, where numerical pre-factors  are omitted. Stability requires $\delta E > 0 $, implying that $k/l^*{}^2>p$, or $\delta z> (p/k)^{1/2}$. As we shall see below, for repulsive particles the bulk modulus $B$ simply follows $B\sim k$, and we can rewrite our result as:
\be
\label{3}
\delta z> (p/B)^{1/2}
\ee
Physically, this results signifies that  pressure has a destabilizing effect, which needs to be counterbalanced by the creation of more contacts to maintain elastic stability. The result is more general  \cite{matthieu2} and for sphere packing one also gets $\delta z\equiv z-z_c>(p/B)^{1/2}$.   $p/B$ is a measure of the contact strain, and for all interaction potential  of interest (approximatively harmonic for emulsion or hertzian for elastic body), one finds that $p/B\sim (\phi-\phi_c)$, where $\phi_c$ is the packing fraction at the jamming threshold (this comes from the fact that $B\equiv \partial p/\partial \phi$, leading to $p/B\sim (\phi-\phi_c)$ if power-law behavior are assumed), so that Eq.(\ref{3}) can be rewritten $\delta z > (\phi-\phi_c)^{1/2}$.
Packings generated numerically  \cite{durian,J} are consistent with the equality of this inequality, supporting that the packing lie indeed close to marginal stability: these systems have  just enough contacts to counterbalance the destabilizing effect of compression.

\section{Elastic Moduli}
\label{s2}

\begin{figure}[htbp]
\begin{center}
\rotatebox{0}{\resizebox{4.5cm}{!}{\includegraphics{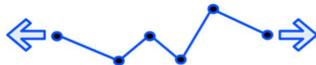}}}
\caption{Zig-zag chain of springs. When pulled at the tips, such a chain yields without stiffness, until the line is straight and forces appears in the contacts. For larger strain the stiffness jumps to a finite value. 
More generally weakly-coordinated networks (with a moderate $\delta z$) are soft when an infinitesimal strain is imposed in a  generic direction, but are stiff if this direction corresponds to the stress (called sometimes pre-stress) sustained by the system. }
\label{f2}
\end{center}
\vspace{-0.5cm}
\end{figure}

Thus, an assembly of repulsive particles close to the jamming threshold presents a vanishing excess-coordination with respect to the jamming threshold: $\delta z \rightarrow 0$ as the strain $p/B \rightarrow 0$.
Why the proximity of the Maxwell bound should impose that shearing becomes much softer than compressing is not obvious a priori. Our argument  will show that  this behavior is specific to purely repulsive systems, and that it is  in fact possible to create spring networks where the behavior is opposite, for which $B/G \rightarrow 0$.  More generally, we shall see that weakly-connected systems (i.e. $z$ close to $z_c$) are  soft to most imposed strains, for which they display a dimensionless stiffness or elastic constant of order $\delta z$ (or zero if the system is floppy, i.e. $\delta z<0$),  with one exception: if the infinitesimal strain is imposed in the direction of the  stress sustained by the system in its reference configuration, the ``pre-stress". In this case the response of the system is not soft and the elastic constant found is similar to the one of a well-connected solid. A simple example is given by the zig-zag chain of springs of Fig(\ref{f2}). When no forces are applied, the chain is in a zig-zag configuration.  When pulled by the two ends, it yields freely up to the point where the line becomes straight and contact forces appear in the springs. At that point, the system does not yield freely anymore, it is  stiff. In what follows we shall show how this simple idea applies to more complex  networks. 

\subsection{Force balance and contact deformation operators}

We are left with determining the elastic moduli of packing of a given coordination. We need to introduce some amount of linear algebra. The argument proposed  is a  shortened version of Ref.\cite{these}. Many of the properties discussed below concern the response of the system to external forces. It proves convenient to consider our system under the influence of an arbitrary set of forces $\vec{F_i}$ acting on all particles $i$. At equilibrium  forces are balanced on each particle $ i$:
\be
\label{72}
\sum_{\langle j \rangle} f_{ij} \vec{n}_{ij}= \vec{F_i}
\ee
where $ f_{ij}$ is the compression in the contact $<ij>$, the sum is on all the particles $j$ in contact with particle $i$, and $\vec{F_i}$ the external force applied on $i$.  $\vec{n}_{ij}$ is the unit vector going from $i$ to $j$. This linear equation can be written:
\be
\label{710}
{\cal T} |{\bf f}\rangle = |{\bf F}\rangle
\ee
$|{\bf f}\rangle$ is  the vector of contact tensions and has $N_c$ components. ${\bf F} $ is the vector of external forces. Its dimension is\footnote[1]{ More precisely, its dimension is $ (N d-d(d+1)/2) $. The term $d(d+1)/2$ corresponds to the constraints on the total torques and forces that must be zero at equilibrium.} $Nd$, since there are $d$ degrees of freedom for the external force on each particle. Therefore ${\cal T} $ is an $ Nd \times N_c$ matrix. In what follows our notations shall be lower-case for the contact space of dimension $N_c$, and upper-case for the particles positions space of dimension $Nd$.

Another important linear operator  describes the change of distances  between particles for a given set of displacements $\delta\vec{R_i}$:
\be 
\label{s}
(\delta\vec{R_i}-\delta\vec{R_j}).\vec{n}_{ij}=\delta r_{ij}
\ee
It can be written as:
\be
\label{dis}
{\cal S}|\delta{\bf R}\rangle = |{\bf \delta r}\rangle
\ee
where ${\bf \delta r}\equiv \{\delta r_{ij}\}$ is the set of distance changes for all contacts.  ${\cal S}$ is an $Nd$ by $N_c$ matrix \footnote[2]{If one removes the global translations or rotations from the displacement fields, which obviously do not change any inter-particle distance, ${\cal S}$ becomes a $(Nd-d(d+1)/2)$ by $N_c$ matrix.}.

As was shown e.g. in \cite{Tom2,roux}, ${\cal T}$ and ${\cal S}$ are transpose of each other, as we now prove. At equilibrium, any force field applied should cost no energy at first order. 
This is the virtual work theorem, which reads here:
\ba
\label{73}
\sum_{i} \delta\vec{R_i}\cdot \vec{F_i}- \sum_{ij} \delta r_{ij}  f_{ij}\equiv \langle\delta {\bf R}| {\bf F}\rangle - \langle \delta {\bf r} | {\bf f}\rangle=0
\ea
where we used the scalar product notation notation $ \langle\delta {\bf R}| {\bf F}\rangle\equiv  \sum_{i} \delta\vec{R_i}\cdot \vec{F_i}$  and $\langle \delta {\bf r} | {\bf f}\rangle\equiv \sum_{ij} \delta r_{ij}  f_{ij}$. Applying the definitions of ${\cal S}$ and ${\cal T}$ in (\ref{73}) we get:
\be
\langle {\bf f}| {\cal S}|\delta {\bf R}\rangle= \langle\delta {\bf R}| {\cal T}|{\bf f}\rangle 
\ee
Introducing the transpose notation this is equivalent to:
\be
{\cal S=T}^t
\ee

\subsection{Energy expansion and virtual force field} 

For simplicity of notation we shall think about rather mono-disperse emulsions, where the interaction potential is approximatively harmonic, of stiffness $k$.
The energy expansion follows Eq.(2) \footnote[30]{When forces are present in the contact, as is the case at finite pressure, another term enters the energy expansion, see  \cite{matthieu2}. It does not affect the scaling of the elastic moduli, and we shall ignore it.}, which can be rewritten:
\be
\label{7M}
\delta E=k\sum_{<ij>} \frac{1}{2} \delta r_{ij}^2=\frac{k}{2}\langle {\cal S} \delta{\bf R}| {\cal S}\delta {\bf R}\rangle 
\ee 
Our system is equivalent to a set of point particles interacting with springs. In order to study its elasticity, it turns out to be convenient to consider the responses that follow arbitrary changes of rest length of these springs. This is in fact equivalent to imposing dipoles of force. As we shall see, the response to shear or compression can also be  easily expressed in terms of changes of rest length.
We impose an infinitesimal change of rest length on every contact ${\bf y}=\{y_{ij}\}$. The energy and the displacement field are given by the minimization of:
\be
\label{74}
\delta E= \frac{k}{2} \min _{\{\delta {\bf R}\}} (\delta r_{ij}-y_{ij})^2= \frac{k}{2} \min _{\{\delta {\bf R}\}} \langle{\cal S}\delta {\bf R}-{\bf y}| {\cal S}\delta {\bf R}-{\bf y}\rangle
\ee
Obviously if ${\cal S}$ was spanning its image space, we would have $\delta E=0$ : one could always find a displacement $|\delta {\bf R}\rangle$ that leads to a change of distances between particles in contact exactly equal to $|{\bf y}\rangle$.  As we said,  ${\cal S}$ is a $N_c\times Nd$ matrix. If $N_c< Nd$,   ${\cal S}$ indeed spans its image space, and the energy associated with any strain $|{\bf y}\rangle$ is zero: the system is floppy.  In the other case, if $N_c> Nd$, there are $ N_c- Nd\equiv N\delta z/2$ relations of dependency among the columns of ${\cal S}$. One can choose a basis of  $ N \delta z/2$ vectors  $|{\bf a}^p\rangle$, with $1\leq p \leq  N\delta z/2$,  in the space of $|\delta {\bf r\rangle}$ such that:
\be
\langle{\bf a}^p| {\cal S}=0
\ee
The $|{\bf a}^p\rangle$ are orthogonal to all the vectors ${\cal S}|\delta {\bf R}\rangle$, for any displacement field  $|\delta {\bf R}\rangle$.  Transposing this relation we have:
\be
{\cal T} |{\bf a}^p\rangle=0
\ee
which indicates that all the vectors in the space of the $|{\bf a}^p\rangle$  satisfy force balance without external force (\ref {72}), but no others. The  $|{\bf a}^p\rangle$  live in the contact-force space, and henceforth we shall denote them $|{\bf a}^p\rangle \equiv |{\bf f}^p\rangle = \{f_{ij}^p\}$. In the following we consider an orthogonal unit basis:
\be
\langle{\bf f}^p| {\bf f}^{p'}\rangle \equiv \sum_{ij} f_{ij}^p f_{ij}^{p'}= \delta_{pp'}
\ee
We can decompose any $|{\bf y}\rangle$ as:
\be
\label{yo}
{\bf y}={\bf y}^{\bot}+\sum_{p=1...N\frac{\delta z}{2}} \langle{\bf f}^p|{\bf y}\rangle {\bf f}^p
\ee
${\bf y}^{\bot}$, the part of $|{\bf y}\rangle$ orthogonal to the $|{\bf f}^p\rangle$, is spanned by the matrix ${\cal S}$, and therefore does not contribute to the energy when the minimization of Eq.(\ref{74}) is performed.  In other words, there exists a displacement  field  $\delta {\bf R_0}$ which leads to a strain ${\bf y}^{\bot}={\cal S}\delta {\bf R_0}$. On the other hand, the strain field corresponding to the second term in Eq.(\ref{yo}) is orthogonal to the space generated by ${\cal S}$, and cannot be accommodated by any displacements. The minimum of Eq.(\ref{74}) therefore occurs in $\delta {\bf R_0}$ and one finds for the energy:
\ba
\label{rere}
\delta E=\frac{k}{2} \sum_{p=1,..\frac{N\delta z}{2}} \langle{\bf f}^p|{\bf y}\rangle^2
\ea
We now discuss properties of the contact force fields that we shall use to derive the scaling of the bulk and the shear moduli. 
Only one vector of the  vector space of the force fields $|{\bf f}^p\rangle$ solutions of Eq.(\ref{710}) without external force is the real set of contact forces  that supports the system, and that could be observed empirically.  This vector is denoted $|{\bf f}^1\rangle$. The rest of the basis  $|{\bf f}^p\rangle$ with $p \neq 1$ are also solutions of  Eq. (\ref{710}) without external force. Nevertheless, there are not ``physical'' solutions for the interaction potential chosen.  Thus we shall call them $\it virtual$. $|{\bf f}^1\rangle$ verifies the following properties: (i) In a system with repulsive interaction,  as we consider here, all the contact forces are compressive and therefore  $ f_{ij}^1>0$ for all contacts. (ii) It is well known from simulations and experiments that the distribution of contact forces is roughly  exponential, or compressed exponential (see for example \cite{J} for simulations in the frictionless case). This implies that the fluctuations of the contact forces are of order of the average value, leading to $\langle f_{ij}^1\rangle^2\sim \langle (f_{ij}^1)^2\rangle=1/N_c$ for a normalized force field.  Thus we may introduce a constant $c_0$ such that:
\be
\label{coco}
\langle f_{ij}^1 \rangle = c_0 \frac{1}{\sqrt {N_c}}
\ee
Now we turn to the properties of  the virtual forces $|{\bf f}^p\rangle$: (i) There are no physical constraints on the sign of the contacts force for the virtual vectors. Furthermore, the ${|\bf f}^p\rangle$ must be orthogonal to  $|{\bf f}^1\rangle$, whose signs of contact forces are strictly positive, and where the fluctuations in the contact compression is small. Therefore  the virtual force fields have  roughly as many compressive as tensile contacts.

\subsection{Elastic moduli}

In our framework it turns out to be convenient to study the response to shear or compression as there are generally implemented in simulations. When periodic boundaries conditions are used, an affine strain is first imposed on the system. Then the particles are let to relax. In general the affine strain is obtained by changing the boundary condition. Consider a 2-dimensional system with periodic boundary: it is a torus. For example a shear strain can be implemented by increasing one of the principal radii  of the torus and decreasing the other. Then the distance between particles in contact increases or decreases depending on the direction of the contact. In fact, this procedure of change of boundary conditions is formally equivalent to a change of the metric of the system. If the metric is changed from identity $ I$  to the constant metric $ G= I+U+U^t $, where $U$ is the imposed infinitesimal global strain, the length of a vector $\vec{\delta l_0}$ becomes $\delta l$, such that $\delta l^2= \vec{\delta l_0} \cdot G \cdot \vec{\delta l_0}$.  Using this expression with $\vec{\delta l_0}=\vec{R_{ij}}\equiv \vec{R_j}-\vec{R_i}$ one deduces the change of distance between two particles: 
\ba
\delta r_{ij}=\vec{n}_{ij}\cdot \frac{U+U^t}{2} \cdot \vec{R_{ij}}
\ea
Near jamming, for mono-disperse particle of diameter $l_0$, $\vec{R_{ij}}\approx l_0 \vec{n}_{ij}$ and therefore $\delta r_{ij}\approx l_0 \vec{n}_{ij}\cdot (U+U^t) \cdot \vec{n}_{ij}/2$. Formally,  such a change of metric is strictly equivalent to a change of the rest length of the springs with $y_{ij}=\delta r_{ij}$. Incidentally  Eq.(\ref{rere}) can be used to compute the energy of such strain.

For a compression $(U+U^t)/2=-\epsilon  I$ where $ I $ is the identity matrix and $\epsilon$ is the magnitude of the strain. Eq.(\ref{rere}) becomes:
\be
\label{75}
\frac{\delta E}{kl_0^2}=\frac{1}{2} (\sum_{ij}-\epsilon f_{ij}^1)^2 + \frac{1}{2} \sum_{p=2,..\frac{N\delta z}{2}} (-\epsilon \sum_{ij} f_{ij}^p )^2 
\ee
In the first sum all the terms have the same sign for a purely repulsive system, and this term leads to the strongest contribution. We have:
\be
\frac{\delta E}{kl_0^2}\geq (\sum_{ij}-\epsilon f_{ij}^1)^2 = \epsilon^2 (N_c \langle f_{ij}\rangle)^2= \epsilon^2 c_0^2 N_c  
\ee
On the other hand, $\delta E$ is certainly smaller than an affine compression  whose energy also goes as $k \epsilon^2 N$.  Therefore we find that:
\ba
\label{12}
\delta E \sim k N \epsilon^2 l_0^2\\
B \equiv \frac{\delta E}{V \epsilon ^2}\sim k l_0^{d-2}
\ea
which does not depend on $\delta z$. Here $V\sim N l_0^d$ is the volume of the system. The bulk modulus of an harmonic system  jumps from 0 in the fluid phase toward a constant when the system becomes jammed, as observed in the simulations. From this result follows that $p\sim B(\phi-\phi_c)\sim(\phi-\phi_c) $. Note that this result holds only for purely repulsive systems. If the potential has an attractive component, and if the pressure is set to zero, the real force field $|{\bf f}^1\rangle$ presents as many negative contact forces as positive ones, and this term does not lead to a particularly large contribution in Eq.(\ref{75}). In this case, one expects generically to  recover for the bulk modulus the result valid for the shear modulus, that we derive in the next section.

If a pure a shear strain is imposed, the tensor $(U+U^t)/2$ is  traceless. Let be $\epsilon$ the largest eigenvalue (in absolute value). The change of distance of two particles in contact due to a pure shear $\delta r_{ij}$ is a number of zero average if the system is isotropic, and fluctuates between  $+\epsilon$ and $-\epsilon$ depending on the orientation of $\vec{R}_{ij}$.  Eq.(\ref{rere}) becomes:
\be
\label{rrr}
\delta E= \frac{k}{2} \sum_{p=1,..\frac{N\delta z}{2}} ( \sum_{ij} f_{ij}^p \delta r_{ij})^2 
\ee
Each term in the summation gives on average:
\ba
\label{pg}
\langle(\sum_{ij} f_{ij}^p \delta r_{ij})^2\rangle = \sum_{ij} \langle (f_{ij}^p)^2 \delta r_{ij}^2\rangle+\sum_{mn\neq ij}  \langle f_{ij}^p f_{mn}^p\delta r_{ij} \delta r_{mn}\rangle \\
= \sum_{ij} \langle(f_{ij}^p)^2\rangle\langle \delta r_{ij}^2\rangle+\sum_{mn\neq ij} \langle f_{ij}^p f_{mn}^p\delta r_{ij} \delta r_{mn}\rangle
\ea
 For a generic imposed strain, $\delta r_{ij}$ and $ \delta r_{mn}$ are not correlated, so that the terms $\langle f_{ij}^p f_{mn}^p\delta r_{ij} \delta r_{mn}\rangle$ vanish in Eq.\ref{pg}. For a pure shear, spatial correlations between  $\delta r_{ij}$ and $ \delta r_{mn}$ exist, and in order to estimate the shear modulus one must make an extra assumption on the nature of the disorder of the network being considered. 
For isotropic random sphere packings one expects the contact network to be strongly disordered and the propagation of forces to be strongly scattered. We shall therefore assume that the contact forces fields only present weak spatial correlations, as is indeed observed numerically \cite{J} (note that it is possible to build networks with reduced disorder where this assumption breaks down, for example in the example shown in Fig.(\ref{f1}) where long straight lines are present in the microscopic structure. If this system is elongated in the direction of these lines, the restoring force will be large independently of the excess coordination $\delta z$).   Following this assumption for packing of particles, we shall neglect the  terms  $\langle f_{ij}^p f_{mn}^p\delta r_{ij} \delta r_{mn}\rangle $ when $mn\neq ij$.  Concerning the diagonal terms, one has $\delta r_{ij}^2 \approx l_0^2\epsilon^2$ while $\sum (f_{ij}^p)^2=1$ by construction. Thus each term in the $p$ summation is of order $l_0^2 \epsilon^2 \cdot 1$, and:
\be
\label{111}
\delta E \sim k N\delta z \epsilon^2 l_0^2
\ee 
Note that this estimation for the energy  will apply for a generic imposed contact strain field $|{\bf y}\rangle$, since in general such a field would only project weakly on the real force field $|{\bf f^1}\rangle$.
Finally Eq(\ref{111}) implies:
\be
\label{xixi}
G \equiv \frac{\delta E}{V \epsilon^2}\sim k \delta z l_0^{2-d}
\ee
Eqs.(\ref{xixi},\ref{12}) lead to the   result $G/B\sim \delta z$. In the context of frictionless particles, one may use Eq.(\ref{3}) to recover the observed result $G/B\sim (p/B)^{1/2}$. 


\section{Summary and Conclusion}

We have shown that weakly-coordinated elastic networks display an exotic elastic behavior: they are very soft and present an elastic constant proportional to $\delta z$ (of order  $\delta z k l_0^{d-2}$, rather than $k l_0^{d-2}$ expected for a usual solid)  for generic deformations, but recover a usual, normally-connected behavior when the strain is imposed along the pre-stress. These results
apply to interacting particles, with the difference that the typical stiffness $k$ now depends on compression. For emulsion the potential is nearly harmonic  and the dependence of $k$ with compression is negligible.
For elastic particles interacting with a Herztian potential,   one gets $k\sim p^{1/3}$. For a hard sphere glass, which lies below jamming  $\phi<\phi_c$,  the interaction is purely entropic, and is studied in Ref. \cite{brito}. 
One finds $k / k_BT \sim (p/ k_BT)^2$.

A consequence of our analysis is that systems made of repulsive particles, which are always under pressure, are stiff  to compression, with respect to the shear modulus that can be tiny if the contact strain is small. 
At a qualitative level this has been observed empirically in emulsions \cite{mason}, glass beads \cite{roux2} and sand \cite{jacob}. Nevertheless as far as scaling exponents are concerned this results has been backed only numerically, without \cite{J,mason} and with friction \cite{makse,roux2}. In order to observe this phenomenon quantitatively  in nature, emulsions and  micro-gel are presumably good candidates. Ideally one would like to have a system where (i) friction is small, as it limits how close one can get from criticality \cite{van} (ii) thermal effects are small near the jamming threshold, for the same reason. This requires to consider sufficiently large particles. (iii) The osmotic pressure is controlled to vary the distance to threshold reliably.  

Such an experimental system may enable to address additional intriguing questions, in particular how amorphous solids yield under  shear.  From our analysis we predict that as a shear strain is imposed, and as the shear stress builds up, the structure must stiffen. For example if the shear stress can be made of the order of the compression near the jamming threshold -which remains to be established- the shear modulus must become of the order of the bulk modulus.  Equivalently, as force chains orientate to hold the shear stress, the system becomes much stiffer in the directions where force chains align. At a qualitative level such a stiffening has been observed numerically \cite{roux3} with elastic particles, and we have argued that this effect occurs more generally  in various systems such as in gels of semi-flexible polymers \cite{maha}. The small value of the shear modulus reflects the presence of soft collective modes in the amorphous structure \cite{maha}, which stop coupling to the applied strain when stiffening occurs.  It would be interesting to investigate if these modes, and the stiffening they generate, are causally related to the yielding of the structure. The scaling of the maximal strain with compression before yielding would be very informative as well to address this issue.

\section{Acknowledgment}

It is a pleasure to thanks W. G. Ellenbroek, A. Kabla, S. Nagel, V. Vitelli and T. Witten for  helpful discussions.


\begin{thebibliography}{99}

\bibitem{mason} T.G. Mason, J. Bibette and D.A. Weitz, {\it Phys. Rev. Lett.}, {\bf  75}, 2051 (1995); T.G. Mason; M.D. Lacasse; G.S. Grest, et al.
Phys. Rev. E, {\bf  56}, 3150-3166   (1997) 

\bibitem{J} C.S O'Hern, L.E Silbert, A. J. Liu and S.R. Nagel,  Phys. Rev. E  {\bf  68}, 011306 (2003),{\bf 76}

\bibitem{shlomon} S.Alexander,  Phys. Rep.,{\bf  296}, 65 (1998)

\bibitem{Tom1} A.V. Tkachenko and T.A Witten,  Phys. Rev. E, {\bf  60}, 687 (1999);

\bibitem{Tom2} A.V. Tkachenko and T.A Witten,  Phys. Rev. E, {\bf  62}, 2510, (2000)

\bibitem{moukarzel} C.F. Moukarzel,  Phys. Rev. Lett.,  {\bf  81}, 1634 (1998)

\bibitem{durian} D.J. Durian,  Phys. Rev. Lett., {\bf  75}, 4780 (1995)

\bibitem{wouter}W. G. Ellenbroek, E.K Somfai, M. van Hecke, and W. van Saarloos, Phys. Rev. Lett. {\bf 97}, 258001 (2006)

\bibitem{matthieu1}  M. Wyart, S.R. Nagel, T.A. Witten, Europhys. Lett., {\bf 72}, 486-492, (2005)

\bibitem{these} M. Wyart, Annales de Physiques Fr., Chapter 7, {\bf 30}, 1, 2005, or arXiv 0512155
\bibitem{ning}  N. Xu, M. Wyart, A. J. Liu, S. R. Nagel, Phys. Rev. Lett., {\bf  98}, 175502 (2007)

\bibitem{brito} C. Brito and M. Wyart,  Euro. Phys. Letters, {\bf 76}, 149-155, (2006);   Jour. of Stat. Mech- theory and exp., L08003   (2007)
; arXiv:0903.0148 (2009), to be published in Journ. of Chem. Phys.

\bibitem{max} Maxwell, J.C. , Philos. Mag. {\bf 27}, 294-299 (1864)

\bibitem{matthieu2}  M. Wyart, L.E.Silbert, S.R. Nagel, T.A. Witten,  Phys. Rev. E {\bf 72}, 051306 (2005)

\bibitem{roux} J-N Roux,  Phys. Rev. E, {\bf  61}, 6802 (2000)

\bibitem{roux2} I. Agnolin and J-N Roux, Phys. Rev. E, {\bf 76}, 061304   (2007)

\bibitem{jacob} Jacob X, Aleshin V, Tournat V, et al., Phys. Rev. Lett, {\bf 100}, 158003, (2008);  L. Bonneau, B. Andreotti, and E. Clement, Phys. Rev. Lett. {\bf 101}, 118001,   (2008)

 \bibitem{makse}  V. Magnanimo, L. La Ragione, J. T. Jenkins, P. Wangand and H. A. Makse, EPL, {\bf 81} 34006, (2008)
 
\bibitem{van} E. Somfai, M. van Hecke, W. G. Ellenbroek, K. Shundyak, W. van Saarloos , Phys. Rev. E {\bf 75}, 020301(2007). 

\bibitem{roux3} P-E Peyneau, J-N Roux, Phys. Rev. E, {\bf 78} 041307 (2008)

\bibitem{maha} M. Wyart, H. Liang, A. Kabla and L. Mahadevan, Phys. Rev. Lett. {\bf 101}, 215501 (2008)







\end{thebibliography}
\end{document}